\documentclass[aps,preprint,showpacs,showkeys]{revtex4}

\usepackage{graphicx}
\usepackage{amssymb}
\usepackage{bm}

\begin{document}
\setcounter{page}{1}
\newcommand{\re}[1]{(\ref{#1})}
\newcommand{\lab}[1]{\label{#1}}
\newcommand{\ci}[1]{\cite{#1}}
\renewcommand{\baselinestretch}{1.25}
\newcommand{\bfr}{\begin{flushright}}
\newcommand{\bfl}{\begin{flushleft}}
\newcommand{\efl}{\end{flushleft}}
\newcommand{\efr}{\end{flushright}}
\newcommand{\bc}{\begin{center}}
\newcommand{\ec}{\end{center}}
\newcommand{\be}{\begin{equation}}
\newcommand{\ee}{\end{equation}}
\newcommand{\bea}{\begin{eqnarray}}
\newcommand{\eea}{\end{eqnarray}}
\newcommand{\ba}{\begin{array}}
\newcommand{\ea}{\end{array}}
\newcommand{\edc}{\end{document}}
\newcommand{\ul}{\underline}
\newcommand{\ri}{\rightarrow\infty}
\newcommand{\li}{\leftarrow\infty}
\newcommand{\ra}{\rightarrow}
\newcommand{\la}{\leftarrow}
\newcommand{\ds}{\displaystyle}
\newcommand{\dsf}{\displaystyle\frac}
\newcommand{\dt}{\Delta{t}}
\newcommand{\il}{\int\limits}
\newcommand{\pal}{\partial}
\newcommand{\xxx}{{\it{X}}}
\newcommand{\bone}{{\bf 1}}
\newcommand{\gComment}[1]{}
\renewcommand{\gComment}[1]{\textcolor{red}{Gerardo: #1}}

\title[]{Bosonization of 2D Fermions due to Spin and Statistical Magnetic Field
Coupling and Possible Nature of Superconductivity and Pseudogap
Phases Below $E_g$ }
\author{B. \surname{Abdullaev}}
\email[E-mail: ]{babdullaev@nuuz.uzsci.net}
\author{C. -H. \surname{Park}}
\affiliation { Research Center for Dielectric and Advanced Matter
Physics, Department of Physics, Pusan National University, 30
Jangjeon-dong, Geumjeong-gu, Busan 609-735, Korea.}

\date[]{Received 14 February 2006}

\begin{abstract}
A ground state energy variational calculation of anyon gas with
Hamiltonian included the interaction of spins of particles with
anyon vector potential induced, i.e. statistical, magnetic field
exhibits exact cancelation of terms connected with fractional
statistics. This leads to bosonization of anyons due to coupling
of their spins with statistical magnetic field. We presume that at
the dense gas fluctuations of effective spins destroy the coupling
and bosons become anyons. At the assumption that pseudogap (PG)
boundary is temperature independent and when anyons are fermions
we use this model to interpret experimental phase diagrams of
Tallon and Loram hole and electron doped High-$T_c$
superconductors below PG energy $E_g$ and find the qualitative and
quantitative agreement. We do the hypothesis that phase transition
(PT) of bosons into Bose-Einstein condensate is not of second
order, but of first order, close to second one, PG regime is meta
stable phase of bosons, and $E_g=0$ is the critical point of this
PT. Bosons undergo PT into fermions on PG boundary. Described in
the literature non-Fermi quasi-particles might be related to
bosons with effective spins.
\end{abstract}


\maketitle

\section{Introduction}
In 2$D$ systems the concept of anyons provides us with unique
opportunity to introduce antisymmetric property of fermion wave
function into Hamiltonian.  It is assumed that particles are to be
spinless. The background of physics of anyons, the Aharonov-Bohm
effect, itself is interesting phenomenon. Related experimental
investigation \ci{nomata} has been recently performed. Anyons enable
to explicitly investigate relationship of spins of particles and
statistics. From standard courses of relativistic quantum theory
(see, for example, Ref. \ci{berestecky}) it is well known that
particles with integer number of $\hbar /2$ spins should obey a
Fermi statistics. It seems, this connection between spins and
statistics is strong. In this work, however, we show that the gas of
2$D$ anyons or fermions with spin $\hbar /2$ can undergo unexpected
changing of statistics and this occurs due to Zeeman interaction of
spins with statistical magnetic field \ci{ler} produced by vector
potential of anyons. The calculation of ground state energy
expectation value will be carried out in the framework of
variational approach with cut-off parameter regularization
\ci{aormn,aom}, which we developed recently.

\section{Theory and Mathematical Treatment}
The Hamiltonian is \bea \ba{r} \hat
H=\dsf{1}{2M}\ds\sum_{k=1}^N\left[\left(\vec p_k+\vec A_{\nu}(\vec
r_k)\right)^2+M^2\omega_0^2 |\vec{r_k}|^2 \right] \\
+ \dsf{1}{2}\ds\sum_{k=1}^N\left[V(\vec r_k) +
\ds\sum_{j\not=k}^N\dsf{e^2} {|\vec r_{kj}|}\right] \ \lab{gsetup1}
\ea \eea for gas of $N$ anyons with mass $M$ and charge $e$,
confined by $2D$ parabolic well, interacting through Coulomb
repulsion potential, in the presence of uniform positive background
\ci{laupr} $ V(\vec r_k)$. Here $\vec r_k$ and $\vec p_k$ represent
the position and momentum operators of the $k$th anyon in two space
dimensions, \be \vec A_{\nu }(\vec
r_k)=\hbar\nu\ds\sum_{j\not=k}^N\dsf{\vec e_z \times\vec r_{kj}}
{|\vec r_{kj}|^2} \lab{gsetup2} \ee is the anyon gauge vector
potential \ci{wu}, $\vec r_{kj}=\vec r_k-\vec r_j$, and $\vec e_z$
is the unit vector normal to the 2$D$ plane. The factor $\nu$
determines the fractional statistics of  anyon: $\nu=0$ (bosons) and
$\nu=1$ (fermions).

In the variational scheme \ci{aormn} we minimize the expression \be
 E=\dsf{\int \Psi_T^*(\vec R)\hat H \Psi_T(\vec R) \ d\vec R}{\int
\Psi_T^*(\vec R) \Psi_T(\vec R) \ d\vec R} \ . \lab{gsetup3} \ee
Here $\vec R = \{\vec r_1....\vec r_N\}$ is the configuration space
of the $N$ anyons. When energies are expressed in units of
$\hbar\omega_0=\hbar^2/(ML^2)$ and lengths in units of $L$ the
normalized trial wave function in the bosonic representation of
anyons reads \be \Psi_T(\vec
R)=\left(\dsf{\alpha}{\pi}\right)^{N/2}\prod_{k=1}^N
\exp\left(-\alpha\dsf{(x_k^2+y_k^2)}{2}\right). \lab{gsetup7} \ee
Here $ \alpha$ is variational parameter.  The harmonic potential
regularization \ci{aom} with tending number of particles $N$ to
infinity yields the ground state energy of infinite Coulomb anyon
gas.

Now we introduce in the Hamiltonian the term \be \dsf{\hbar
}{M}\ds\sum_{k=1}^N {\hat {\vec s}} \cdot \vec b_k \ , \lab{gsetup8}
\ee with statistical magnetic field \ci{ler} \be \vec b_k = -2\pi
\hbar \nu \vec e_z \ds\sum_{j(k\not =j)} \delta ^{(2)}( \vec r_k-
\vec r_j) \ , \lab{gsetup9} \ee which can be derived if calculates
$\vec b_k = \vec \nabla \times \vec A_{\nu }(\vec r_k)$ by using Eq.
~\re{gsetup2}. The sign in Eq. ~\re{gsetup8} is taken for electrons
with charge $e=-|e|$. For holes, with charge $e=|e|$, we need to
change a sign for $\nu$ in Eqs. \re{gsetup2} and \re{gsetup9}, then
Eq. ~\re{gsetup8}, as also the expectation value for energy, Eq.
~\re{gsetup12}, (see below), will retain the sign.

For $s_z=\hbar /2$ and if we take into account that length unit is
$L$, then $\delta ^{(2)}( \vec r )$ should be replaced by $\delta
^{(2)}( \vec r )/L^2$, hence, \be \dsf{\hbar }{M}\ds\sum_{k=1}^N
{\hat {\vec s}} \cdot \vec b_k=
 -\pi  \nu \dsf{\hbar^2}{ML^2} \ds\sum_{k,j(k\not =j)}
\delta ^{(2)}( \vec r_k- \vec r_j) \ . \lab{gsetup10} \ee

The calculation of expectation value Eq. ~\re{gsetup3} when
Hamiltonian is Eq. ~\re{gsetup10} and wave function $\Psi_T(\vec R)$
is Eq. ~\re{gsetup7} gives (we omit a factor $\hbar^2/(ML^2)$) \bea
\ba{r} -\pi  \nu \ds\sum_{k,j(k\not =j)} \int \Psi_T(\vec R)\ \delta
^{(2)}( \vec r_k- \vec r_j) \ \Psi_T(\vec R) \ d\vec R \\
= \dsf{-\nu \alpha N(N-1)}{2} \ . \lab{gsetup11}  \ea \eea

The total expectation value for energy, Eq. ~\re{gsetup3}, including
all terms of Eq.~\re{gsetup1},  is \be E= \frac{N {\cal N} \ \alpha
}{2} + \frac{N}{2\alpha}+ N{\cal M} \ \alpha^{1/2} - \dsf{\nu \alpha
N(N-1)}{2} \ , \lab{gsetup12} \ee where the term for ${\cal M}$ is
responsible for Coulomb interaction (see Refs. \ci{aormn,aom}).

We did in \ci{aormn} a cut-off parameter regularization and found
${\cal N}=1+\nu(N-1)$. This expression for ${\cal N}$ successfully
describes energy of confined anyons \ci{aormn} with and without
Coulomb interaction as well as one of infinite anyon gas \ci{aom}
with Coulomb interaction. Substituting ${\cal N}$ into energy $E$,
Eq. ~\re{gsetup12}, we see the exact cancelation of terms with $\nu$
factors. This result for energy can be obtained if we put $\nu=0$,
i.e., for case of bosons. As the energy of bosons is lower than one
for anyons and fermions, there appears a coupling of spin with
statistical magnetic field for every particle or bosonization of
2$D$ anyons and fermions.

One can assume the fluctuations of spins coupled to magnetic field.
Therefore, bosons with effective spins might look like as Fermi
particles. However, fermions with different spins are independent
\ci{landaus}. Thus, the spins of bosons interact with each other and
do not interact with spins of another fermions if they exist in the
system. We introduce a some correlation length, inside of which
spins of bosons interact with each other. For temperature $T=0$ we
denote it $\xi_o$. The increase of fluctuations destroys the
coupling, and bosons become the anyons or  fermions. This occurs
when the gain in the energy due to fluctuations of spins of bosons
is equal to energy difference between the anyon (or Fermi) and Bose
ground states.

The interaction of spins of bosons we bring in the form \be
e^{-r_o/\xi_o }  \ds\sum_{k=1}^N {\hat {\vec s}}_{k+\delta} \cdot
{\hat {\vec s_k}} \ . \lab{gsetup13} \ee Here it was introduced a
factor $e^{-r_o/\xi_o }$ with $r_o$ is being the mean distance
between particles. For screened by magnetic field spins $\xi_o$ is
to be assumed phenomenological and taken from experiment.

We establish the explicit form of Eq. ~\re{gsetup13}. The growth of
boson spin fluctuations should cancel term, Eq. \re{gsetup8}, in the
Hamiltonian. Therefore, for dense ($r_o<\xi_o$) Bose gas there
should be ${\hat {\vec s}}_{k+\delta}=-\hbar \vec b_k/M$.

The Hamiltonian of bosonized infinite anyon Coulomb gas with
interaction of spins has a form \bea \ba{r} \hat
H=\dsf{1}{2M}\ds\sum_{k=1}^N\left[\left(\vec p_k+\vec A_{\nu}(\vec
r_k)\right)^2+ M V(\vec r_k)\right] \\
+\dsf{1}{2}\ds\sum_{k,j\not=k}^N\dsf{e^2} {|\vec r_{kj}|}+
\dsf{\hbar (1-e^{-r_o/\xi_o }) }{M}\ds\sum_{k=1}^N {\hat {\vec s}}
\cdot \vec b_k  \ . \lab{gsetup14}  \ea \eea

For anyon Coulomb gas with density parameter $r_s>2$, where $r_s$ is
$r_o$ in Bohr radius $a_B$ units, the approximate ground state
energy \ci{aom} per particle \be \dsf{E_0}{NRy}=-\dsf{c_{WC}^{2/3}
f^{2/3}(\nu,r_s)}{r_s^{4/3}}+ \dsf{7\nu
f^{4/3}(\nu,r_s)}{3c_{WC}^{2/3}r_s^{8/3}} \lab{gsetup15} \ee
expressed in $Ry$, Rydberg energy units, and at $\nu=1$ with $15 \%
$ accuracy represents Quantum Monte Carlo result \ci{tanatar} of
Tanatar and Ceperley  for 2$D$ spin polarized electrons. Here \be
f(\nu,r_s)\approx \nu^{1/2}(1-r_s) e^{-r_s}+
\dsf{c_{BL}^{3/2}r_s/c_{WC}} {1+c_{BL}^{3/2}r_s^{1/2}/c_{WC}} \ .
\lab{gsetup16} \ee In Eqs. \re{gsetup15} and \re{gsetup16}
$c_{WC}=3.2903$ and $c_{BL}=1.2934$.

The expectation value calculation of energy with Hamiltonian, Eq.
~\re{gsetup14}, gives analogous  Eq. ~\re{gsetup15} expression for
gas of bosonized anyons, however, one needs to replace in it $\nu$
by $\nu e^{-r_s/\xi_o }$ and now $\xi_o$ is  in $a_B$ units.

\section{Comparison with experimental phase diagrams}
Microscopical mechanism of High-$T_c$ superconductivity (HTSC), as
also the symmetry of its condensate wave function, are challenging
problems of condensed matter physics. Some aspects of the theory,
related to these topics, have been considered in papers
\ci{askerzade}. We apply the model for clarification of experimental
phase diagrams of hole and electron doped cuprates proposed recently
in Ref. \ci{tallon} and Refs. \ci{onose,zimmers}.

In our treatment for the description of HTCS  we consider the
bosonized  fermions with $\nu=1$. To become fermions bosons should
overcome the energy difference \be \Delta_o^B = \dsf{7 (1-
e^{-r_s/\xi_o }) f^{4/3}(0,r_s)} {3c_{WC}^{2/3}r_s^{8/3}} \ ,
\lab{gsetup17} \ee the gap of superconductivity. Our approach, as in
\ci{aom}, corresponds to spinless or fully spin polarized fermions.
One needs to have deal with normal, i.e., no spin polarized electron
liquid. However, the  accuracy of our calculations is lower than the
difference of Tanatar and Ceperley data for ground state energy for
these both phases of electrons.

We express $\Delta_o^B$ as function of density of dopants
$n_s=1/(\pi r_s^2)$ and connect $n_s$ with $p$, fractional part of
doped hole or electron per atom $Cu$. At  big values of $r_s$ or
small $n_s$ one can neglect the exponential factor in Eq.
\re{gsetup17} and $\Delta_o^B\sim n_s\sim p$. At  small $r_s$ or big
$n_s$, without this factor, $\Delta_o^B$ would have $\Delta_o^B\sim
n_s^{2/3}\sim p^{2/3}$, but $e^{-r_s/\xi_o }$ suppresses this
dependence to zero and the law of it depends from function
$\xi_o(p)$. For this limit of $r_s$ we assume that $\Delta_o^B(p)$
coincides with experimental dependence $E_g(p)$. Extrapolating this
asymptotic expression of $\Delta_o^B(p)$ to small values of $p$ and
equating it to $E_g(p)$ one finds the empirical dependence
$\xi_o(p)$. For it $\xi_o(p)\sim 1/E_g(p)$.

To be sure in the correctness of our 2$D$ density of holes, we
express it in $cm^{-2}$ units and compare with experimental one. For
elementary structural cell \ci{dagotto} of almost all cuprate $ab$
planes $a=3.81 \AA$ and $b=3.89 \AA$. Assuming that it has one atom
of $Cu$, the density is $n_{ab}=N_{ab}\cdot p \  cm^{-2}$, where
$N_{ab}$ is number of elementary cells per $ 1 cm^2$ square. One
obtains $n_{ab}= 6.7472\cdot  p\cdot  10^{14} \ cm^{-2}$. We find
experimental density $n_{ab}^{exp}$ from paper \ci{puchkov} for
Y-123 ($Y Ba_2 Cu_3 O_{7-\delta}$) compound at optimal doping
concentration of holes $p\approx 0.16$. It is $n_{ab}^{exp}=
0.9137\cdot 10^{14} \ cm^{-2}$. Our optimal doping value for
$n_{ab}$ is $1.0795\cdot 10^{14} \ cm^{-2}$. From experiment
\ci{puchkov} also leads important information about approximate
equality of total density of holes to density of HTCS carriers. A
comparison of this value for $n_{ab}$  with experimental
concentration \ci{chang} of carriers for Fractional Quantum Hall
Effect (FQHE) $n_{FQHE}\sim 10^{11} \div 10^{12}\ cm^{-2}$ shows
importance for HTCS, as also for FQHE \ci{laupr}, of long-range, not
screened, Coulomb potential interaction between particles. The
screened potential is supposed to be as justification \ci{anderson}
for the treatments based on the Hubbard model.

\begin{figure}
\begin{center}
\includegraphics[angle=0,width=9.5cm,scale=1.0]{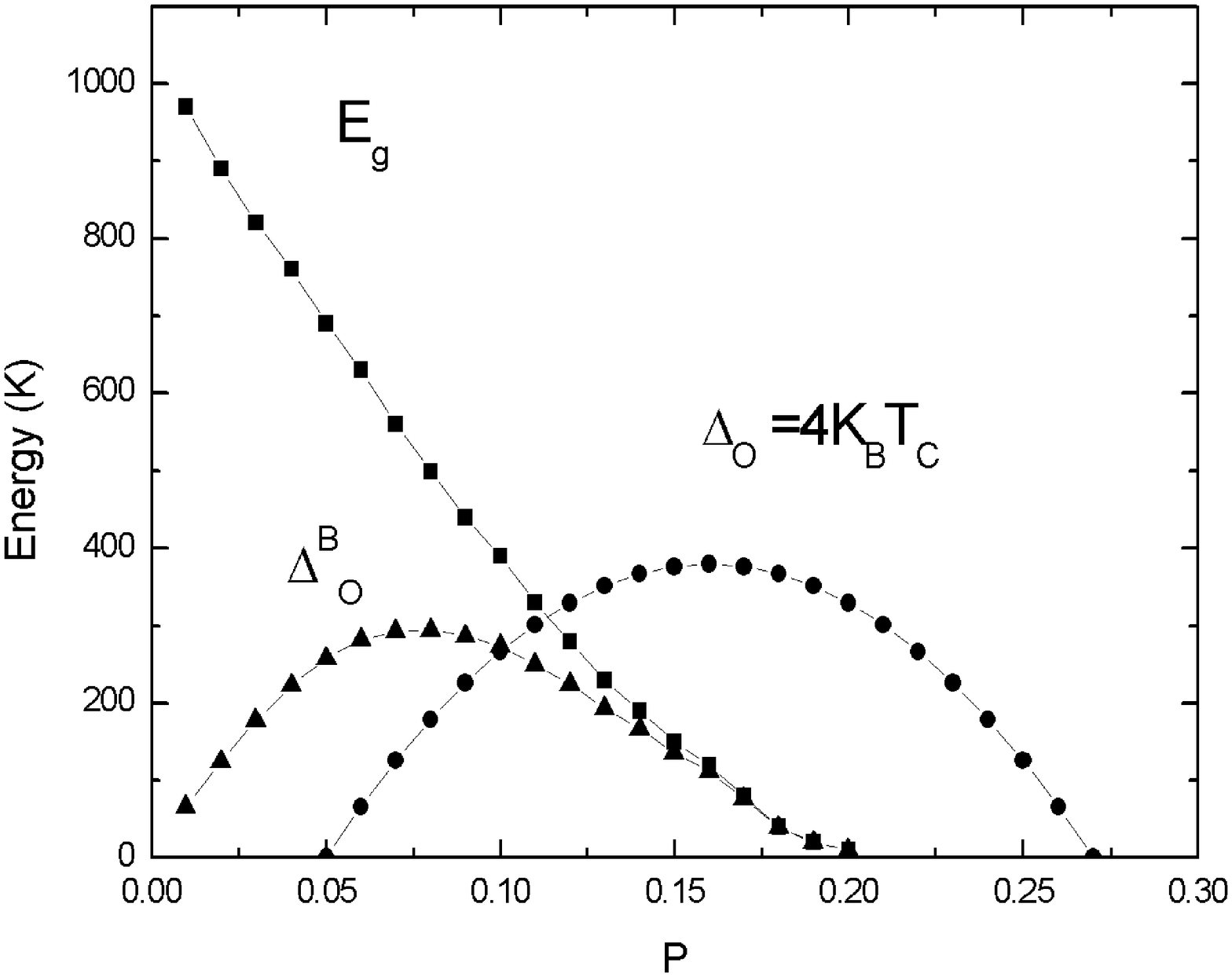}
\end{center}
\caption{ The experimental PG $E_g$, HTCS gap $\Delta_o=4 K_B \ T_c$
(experiment for hole doped $Bi$ \ - 2212 compound), and calculated
from formula Eq. ~\protect \re{gsetup17} HTCS gap for bosons
$\Delta_o^B$ energies in Kelvin temperature (K) units as function of
concentration of holes $p$. } \lab{fig1}
\end{figure}

The Fig. 1 displays  pseudogap (PG) boundary energy $E_g$ (Fig. 11
from paper \ci{tallon}), HTCS gap energy $\Delta_o=4 K_B \ T_c$,
which was evaluated by empirical formula
$T_c=T_{c,max}[1-82.6(p-0.16)^2]$ with $T_{c,max}=95 \ K$ for $Bi $
\ -2212 \ ($Bi_2 Sr_2 Ca Cu_2 O_{8+\delta}$) compound, and HTCS gap
energy calculated from Eq. ~\re{gsetup17} as function of $p$. As we
see our $\Delta_o^B$ has the same magnitude as experimental gap, but
is qualitatively different from generally accepted "dome" like
temperature-concentration phase diagram. However, it is  in
accordance with Fig. 10 of paper \ci{tallon} of Tallon and Loram and
their conclusion that  PG energy $E_g$ up to $p_c\approx 0.19$
separates Bose-Einstein condensate into regions, where density of
Cooper pairs is small and big (weak and strong superconductivity).

\begin{figure}
\begin{center}
\includegraphics[angle=0,width=9.5cm,scale=1.0]{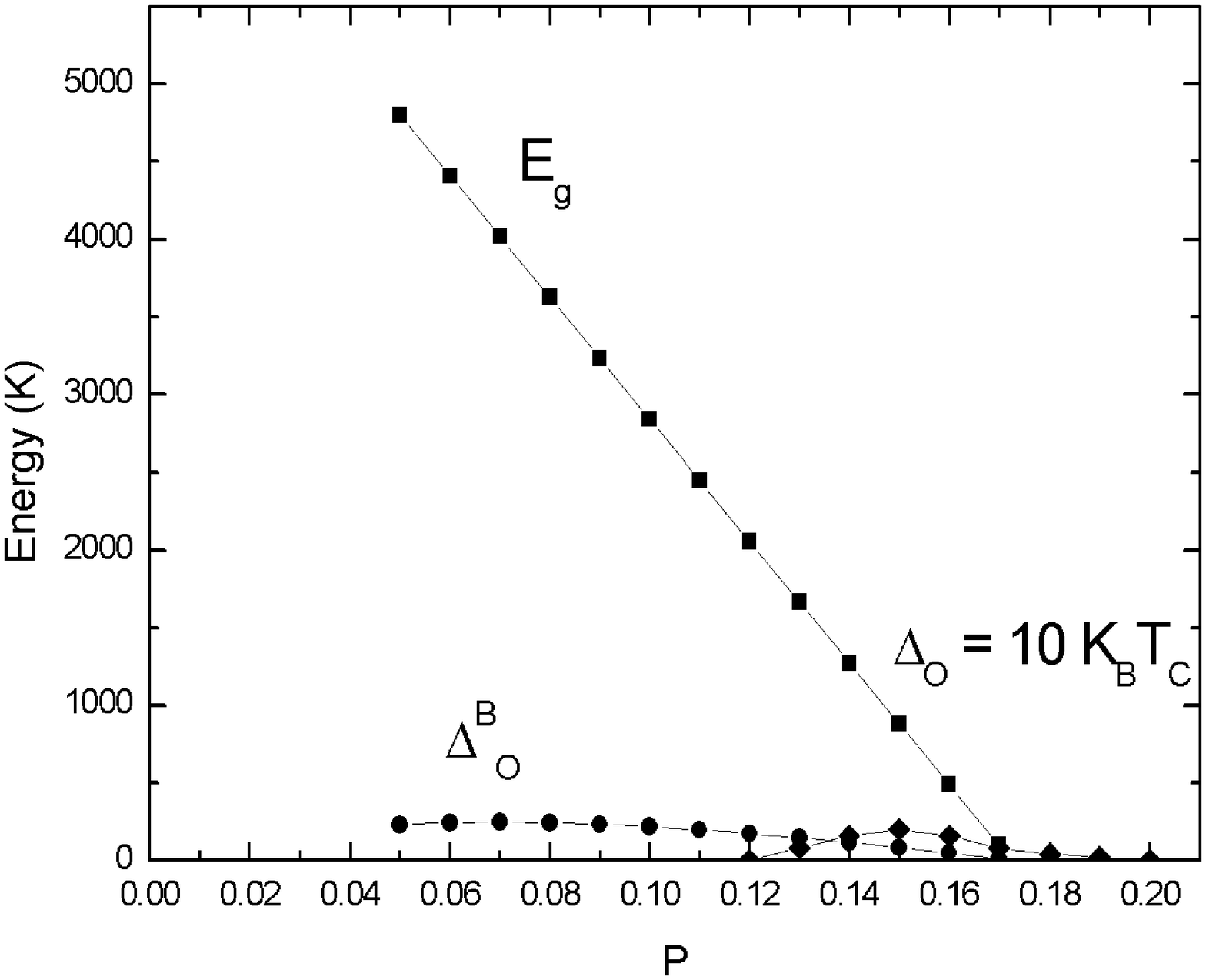}
\end{center}
\caption{ The experimental PG $E_g$, HTCS gap $\Delta_o=10 K_B \
T_c$ (experiments for electron doped NCCO and PCCO compounds), and
calculated from formula Eq. ~\protect \re{gsetup17} HTCS gap for
bosons $\Delta_o^B$ energies in Kelvin temperature (K) units as
function of concentration of electrons $p$. } \lab{fi2}
\end{figure}

For phase diagram data of electron doped cuprates we use Ref.
\ci{onose} for NCCO ($Nd_{2-x} Ce_x Cu O_4$) and Ref. \ci{zimmers}
for PCCO ($Pr_{2-x} Ce_x Cu O_4$). It was shown that $E_g/(k_B
T^*)\approx 10$ for NCCO in  $0.05\le p\le 0.10$ and $E_g/(k_B
T^*)\approx 11$ for PCCO in $0.11\le p\le 0.17$, therefore, we
assume $E_g/(k_B T^*)\approx 10$ for entire interval of $p$ and
interpolate $E_g(p)$ with dependence $E_g=-3916.67p+675.83$. For
experimental HTCS gap we also assume $\Delta_o/(k_B T_c)\approx 10$.
Fig. 2 shows the $p$ dependence of experimental $E_g$, $\Delta_o =
10k_B T_c$ and $\Delta_o^B$ calculated from Eq. ~\re{gsetup17} by
using the above spacing constants of $a$ and $b$ for elementary
structural cell. Comparing with Fig. 1, we see the same qualitative
and quantitative result. More obvious is extension of our
$\Delta_o^B$ to small values of $p$, while experimental $\Delta_o$
starts with $p=0.13$. However, absolute values of both HTCS gaps are
nearly equal.

The spin correlation length $\xi_o$ sharply increases when $p$
approaches $p_c$, which might mean the vicinity of phase transition
(PT). Fig. 6 of Tallon and Loram paper shows that experimental
short-range Anti- Ferromagnetic (AF) correlations scale like
$E_g(p)$ dependence and vanish at $p_c$. The presumption about
correlations of spins of bosons  would mean that there would exist
the competing of these correlations  with AF correlations when
increasing of correlation length $\xi_o$ leads to revealing the
Fermi like spin correlations of bosons in the extended region of
sample and thus, assuming that these spins now interact with ones of
fermions, to suppressing of short-range AF correlations inside of
this region. The definition of competing of HTCS and AF phases,
which is widely used in the literature, in this case would find the
natural interpretation. Only two phases, which have the relative
nature, can compete with each other. AF magnetic phase and based on
Cooper pairs HTCS phase can not coexist, at least, for high, as for
HTCS, temperatures, because they are antagonistic. One would expect
the Cooper pairing in right side from  $p_c$, where the magnetic
correlations are not strong as AF correlations.

We do the hypothesis that due to vicinity \ci{dagotto} of structural
PT to HTCS state in the cuprates for concentration of holes below
optimal the induced by it mechanical strain  would strengthen a
quadratic striction and therefore, the PT into Bose-Einstein
condensate is not a second order, as it should be, but first order,
close to second one \ci{landau2}. The PG regime would correspond to
meta stable phase of bosons. At boundary $E_g$ bosons finally
undergo PT into fermions. The critical concentration $p_c$, at which
$E_g=0$ , would be considered as critical point of first order PT.
In general, the gas of doped holes and electrons in the PG region
might be considered as mixture of Coulomb interacting single
particle bosons and fermions. In the deep underdoped regime and low
temperatures dominate bosons, close to $E_g$ dominate fermions.

\section{Summary}
We have shown that interaction of spins $s_z=\hbar /2$ of anyons
or fermions  with statistical magnetic field leads to
bosonization. For fermions, as the particular case of anyons, we
have applied our model for the possible clarification of phase
diagrams of HTCS of hole and electron doped cuprates for region
below preudogap $E_g$ and found qualitative and quantitative
agreement with experiment.

B. A. gratefully acknowledges  the Korean Research Foundation Grant
(KRF-2004-005-C00044) for support of work.

\end{document}